\documentstyle[amssymb,twoside,fleqn,espcrc2]{article}

\def\@maketitle{
  \global\setbox\fm@box=\vbox\bgroup
    \vskip 2mm
    \raggedright                  
    \hyphenpenalty\@M             
    {\Large \@title \par}         
    \vskip\@bls                   
    {\normalsize                  
     \@author \par}               
    \vskip\@bls                   
    \@address                     
  \egroup
  \twocolumn[
    \unvbox\fm@box                
    \vskip\@bls                   
    \unvbox\abstract@box          
    \vskip 2pc]}                  
\makeatother

\title {Monte Carlo computation of the effective potential
       for the three dimensional Ising system}

\author{Jae -Kwon Kim and D. P. Landau\\
          Center for Simulational Physics, 
          The University of Georgia, Athens, GA 30602}

\begin{document}

\begin{abstract}
Using a novel finite size scaling Monte Carlo technique, we
calculate the four, six and eight point renormalized coupling constants
defined at zero momentum for the three dimensional Ising system. 
Our values of the six and eight point coupling
constants are significantly different from those obtained
from other methods.
\end{abstract}

\maketitle

\section {INTRODUCTION}
Effective action is a useful formalism to describe long distance,
low momentum  behavior of a system. In this work we are interested in
the Monte Carlo calculation of the the effective potential
for the three dimensional (3D) Ising model. 
The Ising model is equivalent to 
the single component $\lambda \phi^{4}$ theory in the limit
$\lambda \to \infty$, so that the use of the standard weak coupling
perturbation method is plagued because of the strong coupling nature
of the model. Hence a variety of new analytical methods have recently
been  developed\cite{BEN1,BEN2,TET,REZ}. 
Due to the non-perturbative nature of the problem, 
Monte Calro methods have also been used. However, standard Monte Carlo
measurements of  relevant physical  
quantities suffer from high statistical noise\cite{WHE}.
In this work we employ a novel finite size scaling (FSS) 
technique\cite{KIM,LAN} combined with a cluster flipping Monte Carlo
algorithm in order to calculate the coefficients of the Taylor expansion
of the effective potential, which are closely related to the
dimensionless N-point renormalized coupling constants (RCC). 

In terms of the renormalized field $\varphi_{R}$ 
the effective potential can be written as
\begin{eqnarray*}
&&V_{eff}(\varphi_{R}) = {1 \over 2} m_{R}^{2} \varphi_{R}^{2}+ 
     {1 \over 4!}m_{R} g_{R}^{(4)} \varphi_{R}^{4}\\
&&\qquad   + {1 \over 6!} g_{R}^{(6)} \varphi_{R}^{6}+
{1 \over 8!} {g_{R}^{(8)} \over m_{R}} \varphi_{R}^{8} + \ldots, 
\end{eqnarray*}
where $m_{R}$ and $g_{R}^{(N)}$ represent respectively the 
renormalized mass (inverse correlation length ($\xi$)) and 
the $N$- point RCC
defined at zero momentum.  It is easy to show that\cite{BEN1}
\begin{eqnarray*}
&&g_{R}^{(4)} = -(Z_{\varphi}^{2}/m_{R})~ W_{2}^{-4}~ W_{4}  \\
&&g_{R}^{(6)} = -Z_{\varphi}^{3}~ W_{2}^{-6}~ [W_{6} - 10 
            W_{4}^{2}~ W_{2}^{-1}] \\
&&g_{R}^{(8)}= -m_{R} Z_{\varphi}^{4}~ W_{2}^{-8}  \\
&& \quad \times [W_{8} - 56W_{6}W_{4}W_{2}^{-1}+280 W_{4}^{3} W_{2}^{-2}]. 
\end{eqnarray*}
Here $Z_{\varphi}$ represents  the field strength 
        renormalized factor ($Z_{\varphi} = \chi m_R^{2}$, 
        with $\chi$ denoting magnetic susceptibility),
	and $W_{N}$ is the Fourier transformed $N$-point
	connected Green function of zero momentum.

\section{THE MONTE CARLO METHOD}
Since we are concerned with the low momentum properties of 
the model when
it is in the scaling regime (but not at criticality), 
our goal is to calculate the thermodynamic values (the infinite volume
limit value) of the RCC measured
at zero momentum in the limit $m_{R} \to 0$.
$g_{R}^{(N)}$ is expressed in terms of the combinations of the
various connected Green functions, so that the statistical noise
in its numerical estimation becomes increasingly large as $N$
increases. 
For a fixed $N$, we empirically observe that the statistical error
increases drastically for approximately $L/\xi \ge 4$. 

\begin{table}[t]
\caption{Size dependence of the various physical quantities at
         $\beta=0.217$.} 
\begin{tabular}{ccccc} \hline
L    &$\xi_{L}$  &$g^{(4)}_{L}$ &$g^{(6)}_{L}$  
&$g^{(8)}_{L}\times 10^{-4}$ \\ \hline
8   &3.93(0)    &9.86(2)  &492.5(1.9) &5.45(3) \\
12  &4.85(1)    &13.59(3)  &852(4)    &10.8(1)\\
16  &5.30(1)    &17.74(6)  &1285(10)  &16.0(3)\\
20  &5.50(1)    &20.95(11) &1591(15)  &16.4(6)\\
24  &5.59(1)    &23.3(1)   &1768(36)  &15.4(1.8)\\
28  &5.60(1)    &24.6(1)   &1958(69)  &18.7(4.7)\\
32  &5.62(1)    &25.1(3)   &1990(187) &18.3(26.4) \\
36  &5.62(1)    &26.0(4)   &2089(278) &        \\ \hline
\end{tabular}
\end{table}

To illustrate we measured various physical quantities 
at an arbitrary (inverse) temperature  with increasing 
linear size of the lattice L.
Table(1) clearly shows that the statistical error 
in the measured values of $g_{R}^{(N)}$ increases with $N$ 
for any given value of L.
For each quantity, the corresponding value increases 
monotonically with $L$ (see also Table(2)) 
until it becomes $L$ independent,
and can be regarded as the corresponding thermodynamic value.
The standard theory of FSS shows that 
the thermodynamic limit
is reached beyond a certain universal value of $L/\xi$, 
{\it regardless of the value of the temperature}.

The correlation length in Table(1) measured 
under the condition $L/\xi_{L} \ge 4$ 
seems to be almost $L$ independent.
However, this is certainly not the case for other 
renormalized quantities which continue to increase significantly
until $L/\xi_{L} \simeq 6$.
Note also that for the six and eight point renormalized couplings
the relative statistical errors increase drastically for larger
values of L, whereas those for the correlation length 
remains negligibly small. 
Evidently, the requirement of larger L for the traditional 
Monte Carlo measurement of the thermodynamic values of the RCC
makes it virtually impossible to measure these quantities precisely
in a deep scaling regime due to the 
large statistical fluctuations.

\begin{table}[t]
\caption{Size dependence of the various physical quantities
         at $\beta=0.2210$ up to $L/\xi_{L} \simeq 3.73$.}
\begin{tabular}{cccccc} \hline
L &$\xi_{L}$ &$g^{(4)}_{L}$&$g^{(6)}_{L}$ &$g_{L}^{(8)}\times 10^{-4}$ 
\\ \hline 
20 & 10.95(2)  &7.6(1)  &301(3)   &2.67(4)  \\
28 & 13.93(6)  &9.2(1)  &423(8)   &4.3(1)  \\
36 & 15.98(6)  &11.2(1) &600(10)  &6.6(2)   \\
48 & 17.9(1)   &14.7(2) &940(20)  &11.3(4)  \\
56 & 18.6(1)   &16.5(4) &1129(60) &13.6(1.2)  \\
64 & 19.0(1)   &18.9(3) &1385(53) &16.2(1.6) \\
72 & 19.3(1)   &20.3(4) &1515(85) &16.3(3.3) \\ \hline
\end{tabular}
\end{table}

We would like to stress that the number of our Monte Carlo
measurements is very large.
For each $L$ we typically generated tens of millions 
of the single cluster configurations. 
For L=28, for example, 
Monte Carlo measurements were taken every 20th configuration out
of a total of $2\times 10^{8}$ configurations that were generated; 
yet, the relative statistical
errors for the $g_{R}^{(6)}$ and $g_{R}^{(8)}$ amount to approximately 
4\% and 25\% respectively. 

In order to overcome the difficulty closer to $T_{c}$,
we make use of a new FSS
function (${\cal{Q}}_{A}(x(L,t))$)\cite{KIM},
defined by the expression
\begin{equation}
A_{L}(t)=A(t){\cal{Q}}_{A}(x(L,t)),~~x(L,t)\equiv \xi_{L}(t)/L.  
\label{eq:fun}
\end{equation}
Here $A_{L}(t)$ represents the quantity $A$ measured on a finite
lattice of linear size $L$ at a reduced temperature $t$,
with its corresponding thermodynamic value $A(t)$ scaling as
$A(t) \sim t^{-\rho}$.

The technique is especially useful for our purpose, because
it enables us to extract accurate thermodynamic values based
on the Monte Carlo measurements on much smaller lattices. 
We just outline the FSS extrapolation technique used in this work. 
For a detailed explanation, we refer the readers to Ref.\cite{LAN}.\\
1).~For a certain $t_{0}$, measure $A_{L}(t_{0})$ and 
       $x(L,t_{0}) = \xi_{L}(t_{0}) /L $ for increasing L.  \\
2).~~Determine $A(t_{0})$ by measuring $A_{L}(t_0)$ 
       such that it is $L$ independent. \\
3).~~Fit $(x(L,t_{0}), A_{L}(t_{0})/A(t_{0}))$ data to a
       suitable functional form. In this work we used the ansatz,
       ${\cal Q}(x) = 1 + c_{1} e^{-1/x} + c_{2} e^{-2/x} + \ldots.$ \\
4).~~For any other $t$, choose a suitable L, measure the value of 
       $x(L,t) \equiv \xi_{L}/L$ and $A_{L}(t)$,
       and interpolate ${\cal Q}(x(L,t))$.  \\
5).~Extract $A(t)$ by plugging $A_{L}(t)$ and ${\cal Q}(x(L,t))$
       into Eq.(1).  \\

Our $t_{0}$ is $\beta=0.217$ and 0.220, and  our largest value of
L for the extraction of the thermodynamic values up to $\beta= 0.2213$
is 72.
\section {RESULT AND DISCUSSION}
\begin{table*}[t]
\setlength{\tabcolsep}{0.80pc}
\caption{Thermodynamic values of the four and six point
         RCC extracted by the 
         FSS technique for some temperatures over
         $0.217 \le \beta \le 0.2213$.}
\begin{tabular} {lccccccr}\hline
$\beta$  &0.217   &0.219  &0.220  &0.2206  &0.2210  &0.2212  &0.2213 \\ 
$g_{R}^{(4)}$ &25.6(4) &25.3(1.1) &24.7(5) &24.6(7) &24.5(3) &24.5(3)&24.4(3)\\
$g_{R}^{(6)}$ &2100(100) &2074(144) &1982(157) &2006(157) &1949(146) 
              &1966(135) &1952(127) \\
      \hline
\end{tabular}
\end{table*}

At $\beta=0.217$ we have  been unable to  determine the precise 
value of the thermodynamic value of $g_{R}^{(6)}$ due to the large
statistical noise in the measurements for L=32 and 36, but
we estimate that $g_{R}^{(6)}(\beta=0.217) \simeq 2100(100)$.
With this value, we have been able to determine the thermodynamic 
values of the $g_{R}^{(6)}$ up to $\beta =0.2213$ 
(where $\xi \simeq 29.2(1)$). 
Of course, all other quantities including correlation length 
have been determined without any ambiguity\cite{LAN} since 
their thermodynamic values  can be accurately measured.
We summarize our results for the RCC
in Table(3), from $\beta=0.217$ to $\beta=0.2213$. 

We observe that $g_{R}^{(4)}$ decreases quite slowly 
as $\xi$ increases, but eventually the value becomes 
stablized around 24.5(2).
$g_{R}^{(6)}$ has a tendency to decrease as well, 
but within the statistical
fluctuations it remains a constant for $\beta \ge 0.2210$, 
i.e., $g_{R}^{(6)} \simeq 1956(145)$.
The statistics of our $g_{R}^{(8)}$ data at $\beta =0.217$
are not good enough for the application of our 
FSS technique. 
At $\beta=0.2210$, however, we have been able to measure 
$g_{R}^{(8)}$ quite precisely (Table(2))
by varying $L/\xi_{L}$  up to 3.73, where we find 
$g_{R}^{(8)} = 1.63(33) \times 10^{5}$.
Again we observe a  systematic increase of $g_{R}^{(8)}$ with 
increasing L, and estimate $g_{R}^{(8)} \simeq 1.75(25) \times 10^{5}$.
Our results indicate that all the RCC ``scale" with the corresponding
values of the critical exponents zero.  
Since the susceptibility and the
correlation length already ``scale" over this regime 
of the temperature\cite{LAN}, such pathological behavior as
a sudden change of them closer to criticality is very unlikely.

Our value of $g_{R}^{(4)}$ is in reasonable agreement with
other estimates, but our estimate of $g_{R}^{(6)}$ is 
more than 40\% larger than those obtained from the measurement
of the probability density of the order parameter\cite{TSY}, 
and from average action method\cite{TET} and 
linked cluster expansion\cite{REZ}.
The disagreement in $g_{R}^{(8)}$ is
more conspicuous; our value is approximately three times larger than
those estimated from the other methods.
The Monte Carlo measurements for the probability distribution 
in Ref.\cite{TSY}, however, were made under the condition 
$L/ \xi_{L} \simeq 4$, which is probably too small.
Our result contradicts the 
claims in the recent literature that the effective potential can
be very well approximated by the terms up to $\varphi^{6}$\cite{TET,TSY}.
On the other hand, our estimates of the renormalized couplings 
are significantly smaller than those estimated 
by the dimensional expansion\cite{BEN2}.
It  has been noticed lately, however, that the higher order terms 
in the expansion become negative\cite{BEN3}.

This research was supported in part by NSF grant number
DMR-9405018.

\end{document}